\documentclass[aps,pre,twocolumn,groupedaddress]{revtex4}
\usepackage{graphicx}
\usepackage{amsmath}
\usepackage{gensymb}

\begin{document}
\title{Dimension of Reservoir Computers}
\author{T. L. Carroll}
\email{Thomas.Carroll@nrl.navy.mil}
\affiliation{US Naval Research Lab, Washington, DC 20375}

\date{\today}

\begin{abstract}
A reservoir computer is a complex dynamical system, often created by coupling nonlinear nodes in a network. The nodes are all driven by a common driving signal. In this work, three dimension estimation methods, false nearest neighbor, covariance and Kaplan-Yorke dimensions, are used to estimate the dimension of the reservoir dynamical system. It is shown that the signals in the reservoir system exist on a relatively low dimensional surface. Changing the spectral radius of the reservoir network can increase the fractal dimension of the reservoir signals, leading to an increase in testing error.
\end{abstract}

\maketitle

{\bf A reservoir computer uses a complex dynamical system to perform computations. The reservoir is often created by coupling together a set of nonlinear nodes. Each node is driven by a common input signal. The time series responses from each node are then used to fit a training signal that is related to the input. The training can take place via a least squares fit, while, the connections between nodes are not altered during training, so training a reservoir computer is fast.

Reservoir computers are described as "high dimensional" dynamical systems because they contain many signals, but the concept of dimension is rarely explored. A reservoir with $M$ nodes defines an $M$ dimensional space, but the actual signals in the reservoir may live on a lower dimensional surface. Two different dimension estimation methods are used to find the dimension of this surface. Counterintuitively, as the dimension of this surface increases, the fits to the training signal become worse. The increase in reservoir dimension can be explained by a well known effect in driven dynamical systems that causes signals in the driven system to have a higher fractal dimension than the driving signal. This increase in fractal dimension leads to worse performance for the reservoir computer. }

\section{Introduction}
A reservoir computer \cite{jaeger2001,natschlaeger2002} is a nonlinear dynamical system that may be used to perform computations on time series signals. Typically the dynamical system is created by connecting a number of nonlinear nodes in a network that includes paths that form cycles, resulting in feedback. Because of the feedback, a reservoir is part of the class of neural networks known as recurrent neural networks.

 An input signal about which one wants to learn something is used to drive the nonlinear nodes. As with other neural networks, before being used as a computer, a reservoir computer must be trained to extract information from the input signal. Unlike other types of neural networks, the network of internal connections for a reservoir computer is not altered during training. Rather, the time series signals produced by the individual nodes are fit to the training signal. As an example, in \cite{lu2017}, the authors sought to reproduce a Lorenz $z$ signal based on a Lorenz $x$ signal, so the $x$ signal was the input and the reservoir computer was trained on the $z$ signal. Because the training can be as simple as a least squares fit, training a reservoir computer can be much faster than training other types of neural network. The output of the training process is a set of coefficients that describe a linear combination of reservoir signals determined by the fitting process.
 
To compute with a reservoir computer, a signal different but related to the original input signal is used to drive the reservoir. In \cite{lu2017}, the new input signal was the $x$ signal from a Lorenz system started with different initial conditions. The new input signal drives the same network as used for training. A linear combination is then made from the new reservoir signals using the coefficients found during the training stage. The signal produced by this linear combination is the output of the computation stage. In \cite{lu2017}, the output of the training stage was the Lorenz $z$ signal corresponding to the new Lorenz $x$ signal.

Because of their simplicity, reservoir computers can be built as analog systems.  Examples of reservoir computers so far include photonic systems \cite{larger2012, van_der_sande2017}, analog circuits \cite{schurmann2004}, mechanical systems \cite{dion2018} and  field programmable gate arrays \cite{canaday2018}. This analog approach means that reservoir computers can potentially be very fast, and yet consume little power, while being small and light. Reservoir computers have been shown to be useful for solving a number of problems, including reconstruction and prediction of chaotic attractors \cite{lu2018,zimmerman2018,antonik2018,lu2017,jaeger2004}, recognizing speech, handwriting or other images \cite{jalavand2018} or controlling robotic systems \cite{lukosevicius2012}.

\subsection{Dimensions}
It is frequently stated that a reservoir computer works by mapping the input signal into a high dimensional space and then projecting this high dimensional space back down to fit the training signal. The entire space is indeed high dimensional, but the reservoir signals may live on a lower dimensional surface within this space. The dimension of this surface is measured here.

Methods for measuring dimension of a dynamical system are just estimates of the actual dimension, and measure different things, so dimensions from different dimension estimation methods will vary. Commonly used methods include the calculation of correlation integrals \cite{grassberger:1983} and false nearest neighbor dimension \cite{kennel1992}. A more recent method which will be used here characterizes embedding dimension by the eigenvalues of covariance matrices \cite{carroll2017a}.

In this work, the embedding dimension of small neighborhoods of points from the full set of signals in reservoir computers is measured. A parameter is varied to change this local dimension, and a correspondence between this local dimension and the performance of the reservoir computer is found. The Lyapunov exponent spectra of the driving signal and the reservoir computer signals are used to estimate the fractal dimension of the individual reservoir signals, and it is shown that changes in this fractal dimension correlate with the reservoir computer performance.

\section{Reservoir Computers}
\label{computers}
\begin{figure}[ht]
\centering
\includegraphics[scale=0.4]{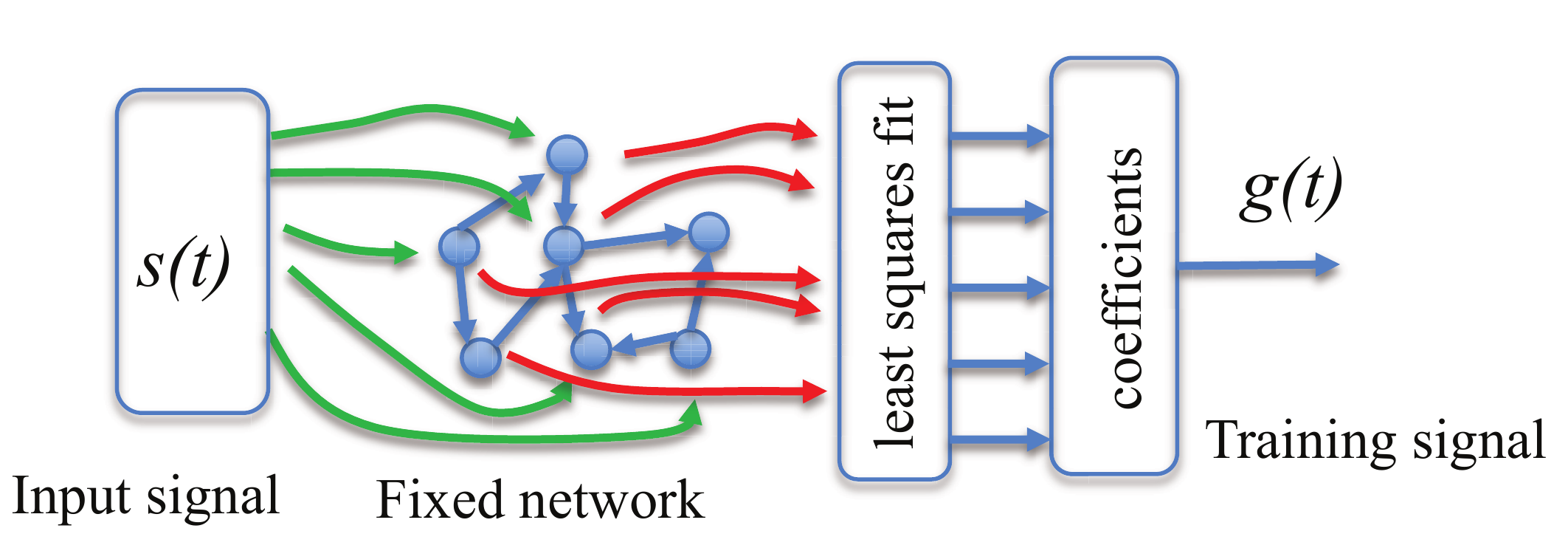} 
  \caption{ \label{reservoir_computer}
Block diagram of a reservoir computer. We have an input signal $s(t)$ that we want to analyze, and a related training signal $g(t)$. In the training phase, the input signal $s(t)$ drives a fixed network of nonlinear nodes, and the time varying signals from the nodes are fit to the training signal $g(t)$ by a least squares fit. The coefficients are the result of the training phase. To use the reservoir computer for computation, a different signal $s'(t)$ is input to the reservoir computer. The time varying node signals that result from $s'(t)$ are multiplied by the coefficients from the training phase to produce the output signal $g'(t)$. Reproduced from Chaos {\bf 29}, 083130 (2019) with the permission of AIP Publishing.}
  \end{figure}

We used a reservoir computer to estimate one time series from a chaotic system based on a different time series from the same chaotic system. Figure \ref{reservoir_computer} is a block diagram of a reservoir computer. There is an input signal $s(t)$ from which the goal is to extract information, and a training signal $g(t)$ which is used to train the reservoir computer. 

A reservoir computer may be described by
\begin{equation}
\label{gen_comp}
{\chi _i}\left( {n + 1} \right) = f\left( {{\chi _i}\left( n \right) + \sum\limits_{j = 1}^M {{A_{ij}}{\chi _j}\left( n \right)}  + {w_i}s\left( t \right)} \right)
\end{equation}

where the reservoir computer variables are the $\chi_i(n), i=1 ... M$ with $M$ the number of nodes, $A$ is an adjacency matrix that described how the different nodes in the network are connected to each other, ${\bf W}=[w_1, w_2, ... w_M]$ describes how the input signal $s(t)$ is coupled into the different nodes, and $f$ is a nonlinear function. 

When the reservoir computer was driven with $s(t)$, the first 1000 time steps were discarded as a transient. The next $N$ time steps from each node were combined in a $N \times (M+1)$ matrix
\begin{equation}
\label{fit_mat}
\Omega  = \left[ {\begin{array}{*{20}{c}}
{{\chi _1}\left( 1 \right)}& \ldots &{{\chi _M}\left( 1 \right)}&1\\
{{\chi _1}\left( 2 \right)}&{}&{{\chi _M}\left( 2 \right)}&1\\
 \vdots &{}& \vdots & \vdots \\
{{\chi _1}\left( N \right)}& \ldots &{{\chi _M}\left( N \right)}&1
\end{array}} \right]
\end{equation}
The last column of $\Omega $ was set to 1 to account for any constant offset in the fit. The training signal is fit by

\begin{equation}
\label{train_fit}
{h(t)} ={\Omega } {{\bf C}}
\end{equation}
where ${h(t)} = \left[ {h\left( 1 \right),h\left( 2 \right) \ldots h\left( N \right)} \right]$ is the fit to the training signal ${g(t)} = \left[ {g\left( 1 \right),g\left( 2 \right) \ldots g\left( N \right)} \right]$ and ${{\bf C}} = \left[ {{c_1},{c_2} \ldots {c_N}} \right]$ is the coefficient vector.

The fit coefficient vector is then found by
\begin{equation}
\label{fit_coeff}
{{\bf C}} = {{ \Omega } _{inv}}{g(t)}
\end{equation}
where $\Omega_{inv}$ is the Moore-Penrose pseudo-inverse of $\Omega$ \cite{penrose1955} and ${{S}}^{{'}}$ is an $(M+1) \times (M+1)$ diagonal matrix constructed from ${ S}$, where the diagonal element $S^{'}_{i,i}=S_{i,i}/(S_{i,i}^2+k^2)$, where $k=1 \times 10^{-5}$ is a small number used for ridge regression \cite{tikhonov1943} to prevent overfitting.

The training error may be computed from
\begin{equation}
\label{train_err}
{\Delta _{RC}} = \frac{{{\rm{std}}\left[ {\Omega {\bf{C}} - g(t)} \right]}}{{{\rm{std}}\left[ {g(t)} \right]}}
\end{equation}
where std[ ] indicates a standard deviation. 

 To learn new information, we use the reservoir computer in the testing configuration. As an example, suppose the input signal $s(t)$ was an $x$ signal from the Lorenz system, and the training signal $g(t)$ was the corresponding $z$ signal. Fitting the Lorenz $z$ signal trains the reservoir computer to reproduce the Lorenz $z$ signal from the Lorenz $x$ signal.

We may now use as an input signal $s'(t)$ the Lorenz signal $x'$, which comes from the Lorenz system with different initial conditions. We want to get the corresponding $z'$ signal. The matrix of signals from the reservoir is now $\Omega'$. The coefficient vector ${\bf C}$ is the same vector we found in the training stage. The testing error is
\begin{equation}
\label{test_err}
{\Delta _{tx}} = \frac{{{\rm{std}}\left[ {\Omega '{\bf{C}} - z'} \right]}}{{{\rm{std}}\left[ {z'} \right]}}
\end{equation}

\subsection{Generalized Synchronization}
If the reservoir computer is repeatedly driven with the same input signal, it must respond in the same way. The reservoir must be in a state of generalized synchronization with the input signal \cite{pecora1990,rulkov1995,kocarev1996,Abarbanel:1996} . The reservoir is finding a function from the input signal to the output signal, which means that there must be a function from the reservoir to the input signal. The probability of having a function from the reservoir to the input signal was measured using a continuity statistic in \cite{carroll2018}.

\section{Node Types}

There are no specific requirements on the nodes in a reservoir computer, other than when all nodes are connected into a network, the network should be stable; that is, when not driven, it should settle into a stable fixed point, and when driven, the same input signal should produce repeatable outputs. Two different node types are used here to decrease the chance that the results depend only on the type of node used.

The polynomial reservoir computer is described by
\begin{equation}
\label{res_comp}
\begin{split}
&\frac{{d{\chi _i}\left( t \right)}}{{dt}} = \\ & \gamma \left[ {{p_1}{\chi _i}\left( t \right) + {p_2}\chi _i^2\left( t \right) + {p_3}\chi _i^3\left( t \right) + \sum\limits_{j = 1}^M {{A_{ij}}{\chi _j}\left( t \right)}  + {w_i}s\left( t \right)} \right].
\end{split}
\end{equation}
The $\chi_i(t)$'s  are node variables, $A$ is an adjacency matrix indicating how the nodes are connected to each other, and ${\bf W}=[w_1, w_2, ... w_M]$ is a vector that describes how the input signal $s(t)$ is coupled to each node. The constant $\gamma$ is a time constant, and there are $M=100$ nodes. This node type will be called the  polynomial  node. For the simulations described here,  $p_1=-3$, $p_2=1$ and $p_3=-1$.  These nonlinear differential equation nodes were chosen because they represent a polynomial, which is a general way to approximate a nonlinear function. The polynomial differential equations were numerically integrated with a 4th order Runge-Kutta integrator. The time step is listed in section \ref{respar} below.

Another node type that was studied was the leaky tanh model from \cite{jaeger2007}
\begin{equation}
\label{umd_comp}
\begin{split}
& {\chi _i}\left( {n + 1} \right) = \\ & \alpha {\chi _i}\left( n \right) + \left( {1 - \alpha } \right)\tanh \left( {\sum\limits_{j = 1}^M {{A_{ij}}{\chi _j}\left( n \right)}  + {w_i}s\left( t \right) + 1} \right).
\end{split}
\end{equation}
This leaky tanh map was also used in \cite{lu2017,lu2018}. The parameters are listed in section \ref{respar}.

\section{Input Vector and Network}
\label{input}
The coupling vector ${\bf W}=[w_1, w_2, ... w_M]$ describes how the input signal $s(t)$ couples into each of the nodes. I want to change only specified parameters in the reservoir computer, so ${\bf W}$ is kept fixed. It has been found that setting all the values of ${\bf W}$ to be the same yields a larger testing error than setting the odd elements of ${\bf W}$ to +1 and the even elements of ${\bf W}$ to -1, so the second method (odd=+1, even=-1) was used. This choice was arbitrary, and other choices of ${\bf W}$ could be made.

The network was kept fixed so that changes in the network would not affect the results of the parameter variation. The effect of different networks on reservoir computer performance was studied in \cite{carroll2019}. The adjacency matrix was initialized to a matrix where 50\% of the network edges were +1 while the others were 0. All the diagonal entries in the adjacency matrix were 0. Of the edges with a value of +1, 50\%  were then flipped to -1. The edges to flip were chosen randomly. The adjacency matrix was then normalized so that the spectral radius $\rho$, defined as the absolute value of the largest real part of the eigenvalues of the matrix, was set to a specified value.

Typically the input vector for a reservoir computer is random and the adjacency matrix is a sparse random matrix. It was shown in \cite{carroll2019} that the input vector and adjacency matrices used here gave the same testing errors as the random versions. 

\section{Input and Training Signals}
The Lorenz system was used to generate input and training signals  \cite{lorenz1963}
\begin{equation}
\label{lorenz}
\begin{array}{l}
\frac{{dx}}{{dt}} = {c_1}y - {c_1}x\\
\frac{{dy}}{{dt}} = x\left( {{c_2} - z} \right) - y\\
\frac{{dz}}{{dt}} = xy - {c_3}z
\end{array}
\end{equation}

with $c_1$=10, $c_2$=28, and $c_3$=8/3. The equations were numerically integrated with a time step of $t_s=0.03$. The training signal was the Lorenz $z$ signal.

In order to test the effect of input signal dimension on reservoir dimension, a nonlinear map of arbitrary dimension was also used to create input and training signals. For an equation dimension $d_e$, the map was described by a matrix $\Phi$:
\begin{equation}
\label{ndmap}
\begin{array}{*{20}{l}}
{{\Phi _{i,j}} = \left\{ {\begin{array}{*{20}{l}}
{\beta \quad i = 1,j = 1}\\
{1\quad i = 1,j = {d_e}}\\
{1\quad i = 2 \ldots {d_e},j = i - 1}\\
{0\quad {\rm{otherwise}}}
\end{array}} \right\}}\\
{{{\bf{x}}_{n + 1}} = \Phi {{\bf{x}}_n}}\\
{{{\bf{x}}_{n + 1}}\left( 1 \right) = {{\bf{x}}_{n + 1}}\left( 1 \right)\quad \,\bmod \,\;1}
\end{array}
\end{equation}.

The constant $\beta=1.1$. The reservoir input signal was $x_1$ and the training signal was the $x_2$ signal. The equation dimension for the map was set to $d_e=3$.

A third system of adjustable dimension was given by the Labyrinth system \cite{thomas2004}. A labyrinth system of equation dimension $d_e$ is given by
\begin{equation}
\label{labyrinth}
\begin{array}{*{20}{l}}
{\frac{{d{x_1}}}{{dt}} = \sin \left( {{x_{{d_e}}}} \right) - b{x_1}}\\
{\frac{{d{x_j}}}{{dt}} = \sin \left( {{x_{j - 1}}} \right) - b{x_j}\quad j = 2 \ldots {d_e}}\\
\quad 
\end{array}.
\end{equation}

The damping constant $b$ was set to 0.2, and the equations were integrated with a 4th order Runge-Kutta routine with a time step of 1. The Labyrinth system is said to follow a chaotic (but not random) walk in phase space. The training signal was the $x_2$ signal. The equation dimension for the Labyrinth system was set to $d_e=10$.

\section{Measurements}
\subsection{Covariance Rank}
\label{comp_rank}
The individual columns of the reservoir matrix $\Omega$ will be used as a basis to fit the training signal $g(t)$. Among other things, the fit will depend on the number of orthogonal columns in $\Omega$.

Principal component analysis \cite{joliffe2011} states that the eigenvectors of the covariance matrix of $\Omega$, $\Theta=\Omega^T\Omega$, form an uncorrelated basis set. The rank of the covariance matrix tells us the number of uncorrelated vectors. 

Therefore, we will use the rank of the covariance matrix of $\Omega$,
\begin{equation}
\label{rank}
\Gamma  = {\rm{rank}}\left( {\Omega ^T\Omega } \right)
\end{equation}
to characterize the reservoir matrix $\Omega$. We calculate the rank using the MATLAB rank() function. The maximum covariance rank is equal to the number of nodes, $M=100$. In \cite{carroll2019}, higher covariance rank was associated with lower testing error.

\subsection{Dimension}
The rank of a set of signals and the dimension of a set of signals need not be the same. An example is a sine wave; the dimension of the sine wave is one, it requires a two dimensional space to be embedded, and its rank is two, because shifting the sine wave by $90\degree$  produces an orthogonal waveform.

The different dimension computation methods described below are measuring different things. The false nearest neighbor dimension \cite{kennel1992} measures how much we know about an attractor in $d+1$ dimensions based on knowing the attractor in $d$ dimensions. The covariance dimension \cite{carroll2017a} estimates the probability, based on the eigenvalues of the covariance matrix, that a $d$ dimensional signal could not have been drawn from a uniform distribution. The false nearest neighbor method characterizes predictability, while the covariance dimension only depends on geometry- the signal being characterized need not be deterministic.

Correlation dimension \cite{grassberger:1983}  is a global measurement that estimates the fractal dimension of a signal. It can be difficult to get good measurements of correlation dimension, especially for high dimensional systems, so correlation dimension is not used here. The Kaplan-Yorke dimension \cite{frederickson1983} is an estimate of the correlation dimension that can be calculated from the Lyapunov exponent spectrum of a chaotic system, so the Kaplan-Yorke dimension is used here in place of the correlation dimension.

\subsubsection{False Nearest Neighbors Dimension $D_{fnn}$}
The false nearest neighbor dimension estimation method was first described in \cite{kennel1992}. To calculate the false nearest neighbor dimension from a single time series, the time series is first embedded in a $d$ dimensional space using the method of delays \cite{abarbanel1993} , and the $n$ nearest neighbors are found. The time series is then embedded in $d+1$ dimensions. The question asked is what fraction of $n$ nearest neighbors in $d$ dimensions are no longer among the $n$ nearest neighbors in $d+1$ dimensions? When the time series requires more than $d$ dimensions for embedding, a large fraction of the nearest neighbors in $d$ dimensions will not be nearest neighbors in $d+1$ dimensions. For some dimension $d$, the false nearest neighbor fraction will drop below some arbitrary threshold. This value of $d$ is taken to be the embedding dimension.

For the reservoir computer, rather than a single time series, there is a time series signal for each node. Instead of embedding a signal in progressively higher dimensions, we create a $d$ dimensional signal by using $d$ individual node signals. From eq. (\ref{fit_mat}), the reservoir computer signals may be arranged in an $N \times (M+1)$ matrix $\Omega$, where $N$ is the number of points in each time series and $M$ is the number of nodes. An $N \times d$ dimensional signal ${\bf u}$ is created by randomly choosing $d$ of the first $M$ columns from $\Omega$. An index point $i_0$ is randomly chosen on ${\bf u}$ and the 10 nearest neighbors to ${\bf u}(i_0)$ are found.

An $N \times (d+1)$ dimensional signal ${\bf u}'$ is then created by adding one more randomly chosen column to ${\bf u}$. The 10 nearest neighbors to ${\bf u}'(i_0)$ are then located. The fraction of false nearest neighbors, $f_{fnn}(i_0,d)$, is the fraction of points that are within the 10 nearest neighbors to ${\bf u}(i_0)$ but not ${\bf u}'(i_0)$.

The fraction of false nearest neighbors upon going from $d$ to $d+1$ is evaluated for 100 randomly chosen values of $i_0$ and the mean value , ${f_{fnn}}\left( {u,d} \right) = {1 \mathord{\left/
 {\vphantom {1 {{N_i}}}} \right.
 \kern-\nulldelimiterspace} {{N_i}}}\sum\limits_{{i_0} = 1}^{{N_i}} {f\left( {{i_0},d} \right)} $, where $N_i=100$  is calculated. Next, a new random set of columns is chosen from $\Omega$ to create a new signal ${\bf u}$ and the process is repeated for a new set of 100 randomly chosen values of $i_0$. Creating the signal ${\bf u}$ is repeated 10 times. There are then 10 values for ${f_{fnn}}\left( {u,d} \right)$, each corresponding to a different random choice of $d$ columns. 

An arbitrary cutoff value must be chosen for the fraction of false nearest neighbors. If the fraction of false nearest neighbors is above this cutoff, it is assumed that going from $d$ to $d+1$ columns reveals more information about the set of reservoir signals . If the fraction of false nearest neighbors is below this cutoff, then adding more than $d$ columns to the signal ${\bf u}$ does not reveal any new information. For this paper, the cutoff was set where the fraction of false neighbors dropped below 0.1. This cutoff is completely arbitrary; the arbitrariness of this cutoff is a problem with the false nearest neighbor method. The false nearest neighbor dimension for each random combination of $d$ columns was $d_{fnn}(u)$. The false nearest neighbor dimension for the reservoir was the mean of the false nearest neighbor dimensions for the $N_u=10$ vectors $\bf u$:
\begin{equation}
\label{dfnn}
{D_{fnn}} = \frac{1}{{{N_u}}}\sum\limits_{k = 1}^{{N_u}} {{d_{fnn}}\left( {{u_k}} \right)} .
\end{equation}

\subsubsection{Covariance Dimension $D_c$}
Dimension may also be estimated by whether a $d$ dimensional signal can be distinguished from a Gaussian random signal with $d$ components. The null hypothesis for this method is that the signal of interest comes from a Gaussian random process, and therefore is isotropic in space. We seek to disprove this null hypothesis by embedding the signal in $d$ dimensions and comparing the eigenvalues of the covariance matrix to what would be obtained for a Gaussian random signal. In \cite{carroll2017a}, the covariance matrix of a delay embedded signal was used to estimate the dimension necessary to embed the signal.

The covariance matrix for the reservoir was already described in eq. (\ref{rank}). In that example, the covariance matrix was calculated for the entire set of reservoir data. In this section, the goal is to detect anisotropy at the smallest length scales in the reservoir, so the covariance matrix is calculated from small clusters of points.

To understand why small clusters of points are necessary, consider the Lorenz system described above. A plot of the Lorenz signals in two dimensions is clearly anisotropic, even though the Lorenz system is three dimensional. Reference \cite{carroll2017a} shows that using by small clusters, the number of dimensions needed to embed the $x$ signal from the Lorenz system is properly calculated as three.

An $M \times d$ dimensional signal $\bf u$ is found by randomly selecting $d$ columns from the matrix $\Omega$, as was done for the false nearest neighbor dimension. A small cluster of $d+1$ points is found by picking an index point $i_0$ and finding the $d+1$ nearest neighbors. Clusters with $d+1$ points are just larger than the smallest clusters useful for calculating a $d$ dimensional covariance matrix.  The cluster of $d+1$ points from  ${\bf u}$ will be called ${\bf v}$, so ${\bf v}$ has dimensions $(d+1) \times d$. This cluster is then normalized by subtracting the mean from each component and dividing by the standard deviation
\begin{equation}
\label{unorm}
{{\bf{w}}_j} = \frac{{{{\bf{v}}_j} - {{{\bf{\bar v}}}_j}}}{{\sqrt {\sum\limits_{j = 1}^d {\sum\limits_{i = 1}^{d + 1} {{{\left[ {{{\bf{v}}_j}\left( i \right) - {{{\bf{\bar v}}}_j}} \right]}^2}} } } }}
\end{equation}
where the overbar operator indicates the mean and the subscript $j$ indicates one of the $d$ components of the vector ${\bf v}$. Next, the $d \times d$ covariance matrix is found
\begin{equation}
\label{covar}
C = \frac{{{{\bf{w}}^T}{\bf{w}}}}{{{d+1}}}
\end{equation}

 A $d$ dimensional Gaussian random process will be isotropic in a $d$ dimensional space. The eigenvalues for a $d \times d$ covariance matrix for a Gaussian random signal $n$ points long are known to converge to a Marchenko-Pastur distribution \cite{marchenko1967} as $n$ and $d$ approach $\infty $. For finite $n$ and $d$, the eigenvalues for the covariance matrix for a Gaussian random signal may be estimated by drawing random $d \times d$ matrices from the Wishart distribution \cite{wishart1928} with a mean covariance matrix equal to the identity. For this paper, the limiting eigenvalues for a Gaussian random process were estimated by creating 10,000 random covariance matrices for each combination of $n$ and $d$ using the MATLAB wishrnd() function.
 
 To study the reservoir signals, a covariance matrix was created from each cluster of points $\bf v$  as in eqs. (\ref{unorm}-\ref{covar}) and the eigenvalues for this matrix were calculated. Once again, 10 random combinations of $d$ columns from $\Omega$ were chosen to create ${\bf u}$, and for each ${\bf u}$ 100 random indices $i_0$ were used to find $d+1$ nearest neighbors.

For a given signal ${\bf u}$, the number of clusters ${\bf v}$ with at least one covariance matrix eigenvalue outside the limits for a Gaussian random process was recorded. A total of 100 clusters were created, and the fraction of clusters with an eigenvalue outside the Gaussian random limits was $f_{ev}$.

 The fraction $f_{ev}$ was calculated for values of $d$ from 2 to 100 for a 100 node reservoir. The lowest value of $d$ for which $f_{ev}$ exceeded 0.9 was taken as the covariance dimension $d_c(u)$ for ${\bf u}$. A fraction of 0.9 meant that there was a 90\% probability that the $d$ dimensional signal ${\bf u}$ was not isotropic, and so did not come from a Gaussian random process. The threshold of 0.9 was arbitrary. The effect of different thresholds for the false nearest neighbor and covariance dimensions was studied in section \ref{lor_dim}.

The covariance dimension for the reservoir was the mean of the covariance dimensions for the individual combinations ${\bf u}$:
\begin{equation}
\label{cdim}
{D_c} = \frac{1}{{{N_u}}}\sum\limits_{k = 1}^{{N_u}} {{d_c}\left( {{u_k}} \right)} .
\end{equation}

\subsection{Kaplan-Yorke Dimension $D_{KY}$}
 The Kaplan-Yorke dimension \cite{frederickson1983} is an estimate of the capacity (or fractal) dimension based on the spectrum of Lyapunov exponents of a chaotic system. For a spectrum of Lyapunov exponents $\lambda_1 \ge \lambda_2 ... \ge \lambda_d$, the Kaplan-Yorke (or Lyapunov) dimension is
\begin{equation}
\label{dimky}
{D_{KY}} = j + \sum\limits_{k = 1}^j {\frac{{{\lambda _k}}}{{\left| {{\lambda _{j + 1}}} \right|}}} 
\end{equation}
where $j$ is the largest integer for which the cumulative sum of the Lyapunov exponents is greater than 0 and the $\left| \; \right|$ operator indicates the absolute value.

The Lyapunov exponents for the driving signals were calculated from the method in \cite{eckmann1985}. For a set of differential equations $\dot x = f\left( x \right)$ a Jacobian $J(t)$ is found and used to propagate a vector $\xi$ that represents a separation between trajectories
\begin{equation}
\begin{array}{l}
J(t) = \frac{\partial }{{\partial x}}f\left( x \right)\\
\dot \xi  = J\left( t \right)\xi 
\end{array}
\end{equation}
To find all $d$ Lyapunov exponents, $\xi$ is a set of $d$ vectors, each of dimension $d$. A commonly used choice for the initial value of $\xi_0$ is the identity matrix. After integrating forward in time, the resulting $\xi_1$ is decomposed with a $QR$ decomposition where $Q$ is an orthogonal matrix and $R$ is an upper triangular matrix. At time step $i$, the diagonal elements of $R$ are $r_j(i)$. To initialize the next time step, $\xi_0=\xi Q$. After many time steps, $\xi$ lies mostly along the direction of the largest Lyapunov exponent, so $\xi$ must be periodically renormalized. The Lyapunov exponents are then calculated from 
\begin{equation}
\label{lyexp}
{\lambda _j} = \frac{1}{{N\tau }}\sum\limits_{i = 1}^N {{{\log }_{10}}} \left( {\left| {{r_j}\left( i \right)} \right|} \right)\quad j = 1 \ldots d
\end{equation}
where the time step is $\tau$ and there are $N$ time steps.

\section{Reservoir Parameters}
\label{respar}
 For the polynomial nodes, value of $\gamma$ was found by minimizing the testing error when predicting the Lorenz $z$ (or $x_2$ from the other two systems) when the input signal was the Lorenz $x$ signal (or the $x_1$ signal from the other two systems), with a spectral radius of $\rho=0.5$. For the Lorenz signal , when both the Lorenz and polynomial equations were numerically integrated with a time step of 0.03 s, the optimum value of $\gamma$ was 4. When the 3-d nonlinear map $x_1$ signal drove the polynomial nodes, with an integration time step of 0.03 for the node equations, the best value of $\gamma$ was 10. When the $x_1$ signal from the 10-d Labyrinth system drove the polynomial nodes, the time step for both Labyrinth and polynomial ode's was set to 1, and $\gamma=0.26$.
  
When the leaky tanh nodes were driven by the Lorenz $x$ signal, the Lorenz integration time step was again 0.03. The optimum value of $\alpha$ was 0.75. When the drive signal was the 3-d map $x_1$ signal, the optimum was $\alpha=0.25$. With the 10-d Labyrinth $x_1$ as a driving signal, with an integration time step of 1, $\alpha=0.68$.

\section{Reservoir Dimension}
\subsection{Input Signal Dimensions}
The dimension of the input signal may influence the reservoir dynamics. The covariance and false nearest neighbor dimensions of the nonlinear map and labyrinth signals were measured as the equation dimension $d_e$ was increased. 

The covariance dimension of the nonlinear map saturated at a value of 5, but the false nearest neighbor dimension of the map was equal to the equation dimension $d_e$. The covariance dimension showed that the map signal did not appear to be a Gaussian random signal for $d_e \ge 5$, but the false nearest neighbor dimension showed that it was not predictable. This difference in dimension measurements could be caused by a finite bandwidth for the map signal; reference \cite{carroll2017a} showed that filtered noise could have a low covariance dimension because it was random but not Gaussian.

The covariance dimension of the Labyrinth signal saturated at 16, while the false nearest neighbor saturated at 37. For this work, $d_e$ for the nonlinear map was set at 3, resulting in a  system with a Kaplan-Yorke dimension of 3. The equation dimension for the Labyrinth signal was set at 10, which caused the Kaplan-Yorke dimension to be 6.43.

\subsection{Reservoir}
As the number of nodes in the reservoir increases, the dimension of the set of reservoir signals should increase. If the nodes were uncoupled and driven by independent signals, we would expect the dimension of the set of reservoir signals to be equal to the number of nodes. The nodes are coupled, however, and they are all driven by the same signal (within a factor of $\pm 1$), so the dimension will probably be less than the number of nodes.

The input signals in all cases were 101,000 points long. The first 1,000 points from each reservoir signal were discarded as a transient.

\subsection{Lorenz System}
\label{lor_dim}
\begin{figure}[ht]
\centering
\includegraphics[scale=0.8]{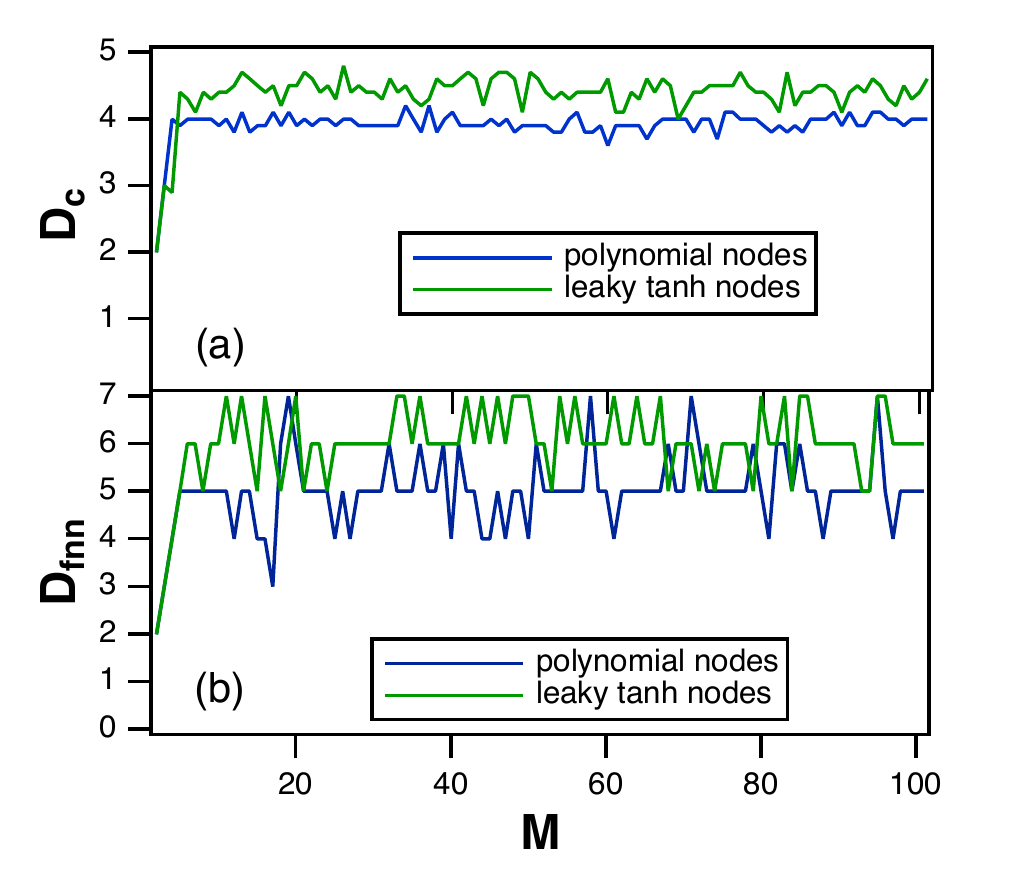} 
  \caption{ \label{lornodes}
(a) Covariance  dimension $D_c$ of the a the two different reservoir types when the input signal comes from the Lorenz $x$ signal and the number of nodes $M$ increases. The leaky tanh nodes give a slightly higher dimension than the polynomial nodes, but the covariance dimension for both node types saturates as the number of nodes increases. (b) False nearest neighbor dimension $D_{fnn}$ as a function of the number of nodes for the same two node types and input signal. Once again, the dimension saturates.}
  \end{figure}

The $x$ signal from the Lorenz had a Kaplan-Yorke dimension of 2.06. The covariance and false nearest neighbor dimensions for the Lorenz $x$ signal were found to be 3. Figure \ref{lornodes} shows the covariance and false nearest neighbor dimensions of the two different reservoir types when driven by the Lorenz $x$ signal.

Figure \ref{lornodes} shows that when the reservoir is driven with the Lorenz $x$ signal, both covariance and false nearest neighbor dimensions saturate at a relatively low value for both node types. The covariance dimension saturates at a mean value of just under 4 for the polynomial nodes and just over 4 for the leaky tanh nodes. The false nearest neighbor dimension saturates at a mean value of about 3.9 for the polynomial nodes and about 5 for the leaky tanh nodes. 

These dimension values all depend on arbitrary thresholds, but the fact that the saturation values are so low definitely indicates that the the reservoir signals occupy a low dimensional surface in a high dimensional space. It is interesting that the dimension of this surface does not increase past a certain value as the number of nodes keeps increasing. 

\begin{figure}[ht]
\centering
\includegraphics[scale=0.8]{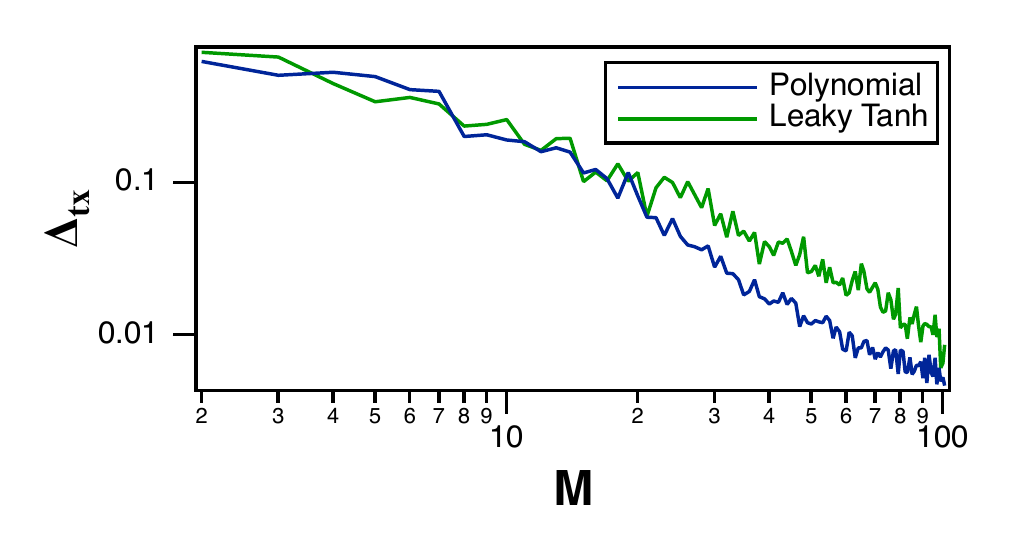} 
  \caption{ \label{lor_nodes_err} Log-log plot of the testing error $\Delta_{tx}$ when the reservoir is trained to predict the Lorenz $x$ signal and the number of nodes $M$ increases.
}
  \end{figure}

Figure \ref{lor_nodes_err} also shows that testing error decreases as the number of nodes decreases even though the dimension of the reservoir signals stops changing above a certain number of nodes.

The effect of different thresholds is not large. If the threshold for the covariance dimension is set to 0.8 and the threshold for the false nearest neighbor dimension is set to 0.2, the covariance dimension for the polynomial nodes saturates at a mean value of about 3.2, while the false nearest neighbor dimension saturates at a mean value of about 3. If the thresholds for the covariance and false nearest neighbor dimensions are set at 0.95 and 0.05, for the polynomial nodes the covariance dimension saturates at a mean value of about 4 while the false nearest neighbor dimension saturates at a mean value of about 8.

As the number of nodes $M$ increases, the testing error $\Delta_{tx}$ decreases as expected. Figure \ref{lor_nodes_err} is a log-log plot of the testing error for both types of nodes driven by the Lorenz $x$ signal. 

The testing error as a function of number of nodes in figure \ref{lor_nodes_err} appears to follow a power law for larger numbers of nodes. The log-log plot of figure \ref{lor_nodes_err} for the polynomial nodes has a slope of approximately -1.66, while for the leaky tanh nodes the slope is -1.35. 

Table \ref{sig_dim} shows the values at which the dimensions saturated for all signals and nodes. The testing error for all combinations of input signal and node type followed a power law as the number of nodes increased; table \ref{sig_dim} also shows the slope for these power laws.

\begin{table}[]
\centering
\caption{Kaplan-Yorke $D_{KY}$  for the different input signals, followed by the covariance dimension $D_c$ and false nearest neighbor dimension $D_{fnn}$ for different 100 node reservoirs driven by the specified input signals. Also shown is the slope of a log-log plot of the testing error $\Delta_{tx}$ as a function of the number of reservoir nodes.}
\label{sig_dim}
\begin{tabular}{|c|c|c|c|c|c|c|}
\hline
input signal  & $D_{KY}$ &  node type &  $D_c$  & $D_{fnn}$   & $\Delta_{tx}$ slope \\
\hline
 Lorenz $x$ & 2.06 &  polynomial & 3.9 & 5 & -1.66\\
 Lorenz $x$ & & leaky tanh & 4.4 & 6.1 & -1.35 \\
 nonlinear map $d_e=3$ $x_1$ & 3.0  & polynomial & 4.8 & 7.3 & -1.08 \\
 nonlinear map $d_e=3$ $x_1$  & &  leaky tanh & 5.5 & 8.7  & -0.73\\
 Labyrinth $d_e=10$ $x_1$ & 6.43  &  polynomial & 4.8 & 6.6  & -0.71\\
 Labyrinth $d_e=10$ $x_1$ &  & leaky tanh & 5.6 & 10.9  & -0.82 \\
 
\hline         
\end{tabular}
\end{table}

Comparing the dimension results for all three signal types, the input signal can influence the dimension of the set of reservoir signals, but the dimension for the reservoir still saturates at a value much less than the total number of nodes. Input signals with larger dimensions were also used to drive the reservoirs, but the reservoir dimensions were the same as in table \ref{sig_dim}. The reservoir dimension appears to be a function of node type and input signal type, but the actual dimension of the input signal, or the number of nodes (above some minimum), appear to have little influence.

The slope of the power law describing how the testing error changes with the number of nodes has a larger magnitude for the polynomial nodes for two of the three driving signal types, indicating that the polynomial nodes are more sensitive to the size of the reservoir computer than the leaky tanh nodes. The magnitude of the slope for the polynomial nodes is slightly smaller than the slope for the leaky tanh nodes when the driving signal is the ten dimensional Labyrinth signal, making it hard to draw any general conclusions about how testing error depends on reservoir size as different nodes are used.

\subsubsection{Effect of Number of Data Points}
Dimension algorithms can give incorrect results if the number of points is insufficient, especially for high dimensional systems. All dimension calculations were repeated for a 100 node reservoir with input signals ranging from 50,000 to 1,000,000 points long (after discarding transients). The dimension results did not change with the number of points.

\section{Changing Reservoir Dimension}
Can the reservoir dimension be changed, and what happens if it does change? A number of reservoir parameters were varied to test their effect on the reservoir dimension, and it was found that the dimension was sensitive to the spectral radius $\rho$. The spectral radius is the largest absolute value of the real part of the eigenvalues of the adjacency matrix $A$.

\begin{figure}[ht]
\centering
\includegraphics[scale=0.8]{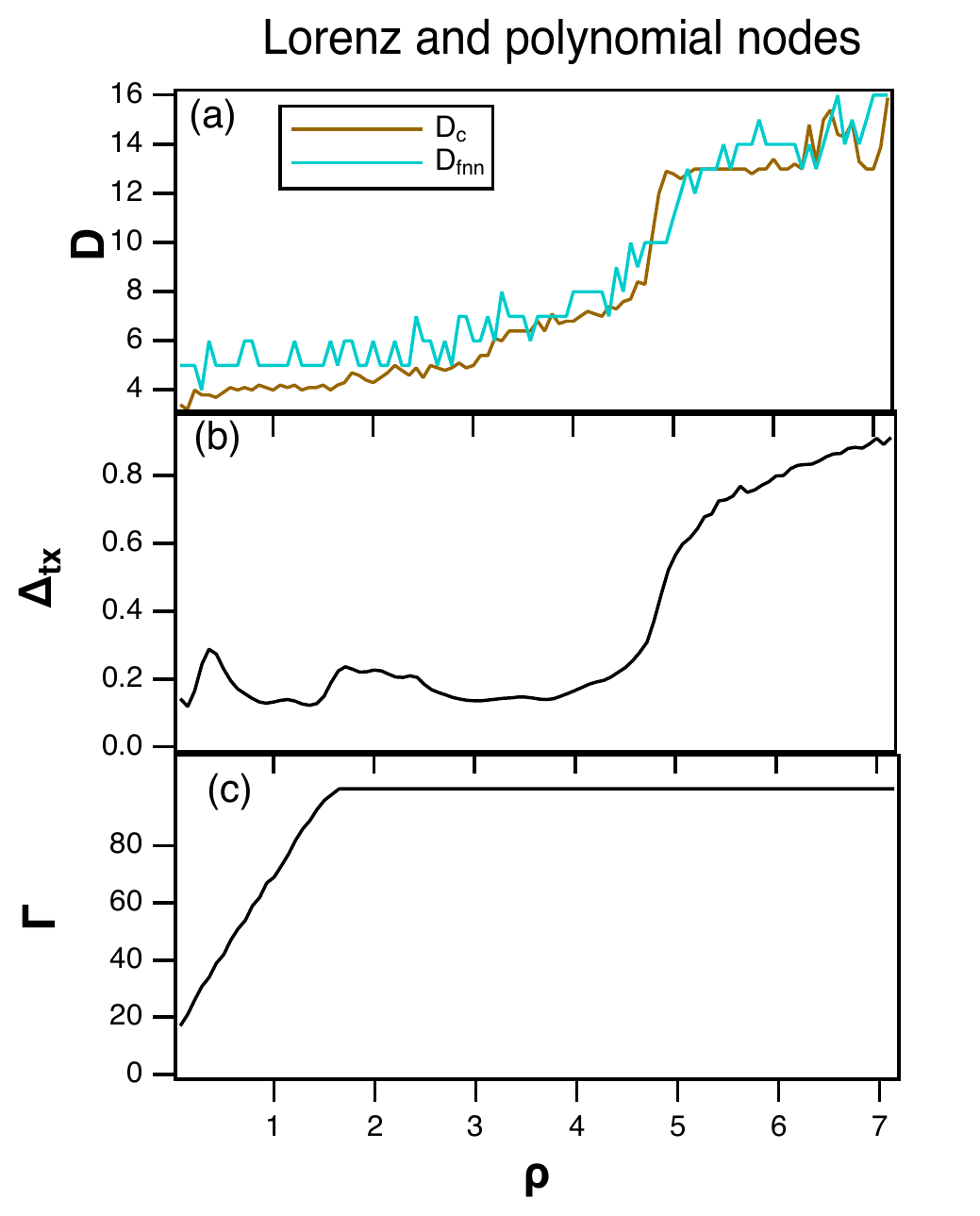} 
  \caption{ \label{lor_poly_dim} (a) Covariance dimension $D_c$ and false nearest neighbor dimension $D_{fnn}$ of a reservoir consisting of 100 polynomial nodes driven by the Lorenz $x$ signal. The parameters were indicated in section \ref{respar}. As the spectral radius $\rho$ increases, both the covariance and false nearest neighbor dimensions increase. (b) Testing error $\Delta_{tx}$ for predicting the Lorenz $z$ signal from the Lorenz $x$ signal. (c) Covariance rank as defined in eq. (\ref{rank}).
}
  \end{figure}

Figure \ref{lor_poly_dim} shows how changing the spectral radius $\rho$ affects the measured dimensions, testing error and the covariance rank of a reservoir consisting of 100 polynomial nodes with the Lorenz $x$ signal as an input. 

  Figure \ref{lor_poly_dim} shows both rank and two types of dimension. It should be noted that rank and dimension are not the same thing. Rank in this case is the number of orthogonal signals that can be created from the set of reservoir signals. Covariance dimension indicates that the set of signals is anisotropic in space if more than $D_c$ signals are combined into a vector. False nearest neighbor dimension describes if a set of $d+1$ reservoir signals can be predicted from a set of $d$ signals. It is not surprising that $D_c$ and $D_{fnn}$ for a reservoir follow similar patterns. For deterministic signals, both types of dimension are related to the number of dimensions needed to embed a signal.

In figure \ref{lor_poly_dim}(a), both covariance and false nearest neighbor dimensions increase as the spectral radius $\rho$ increases. Figure \ref{lor_poly_dim} shows how this dimension increase affects the testing error for predicting the Lorenz $z$ signal from the Lorenz $x$ signal. 

The testing error in figure \ref{lor_poly_dim} changes very little for $\rho < 4$ increases, even though the covariance matrix rank $\Gamma$ is increasing for $\rho < 2$. In \cite{carroll2019} it was shown that the testing error decreased as $\Gamma$ increased. Section \ref{dimchange} explains why the testing error does not decrease, but instead increases for $\rho > 4$.

\begin{figure}[ht]
\centering
\includegraphics[scale=0.8]{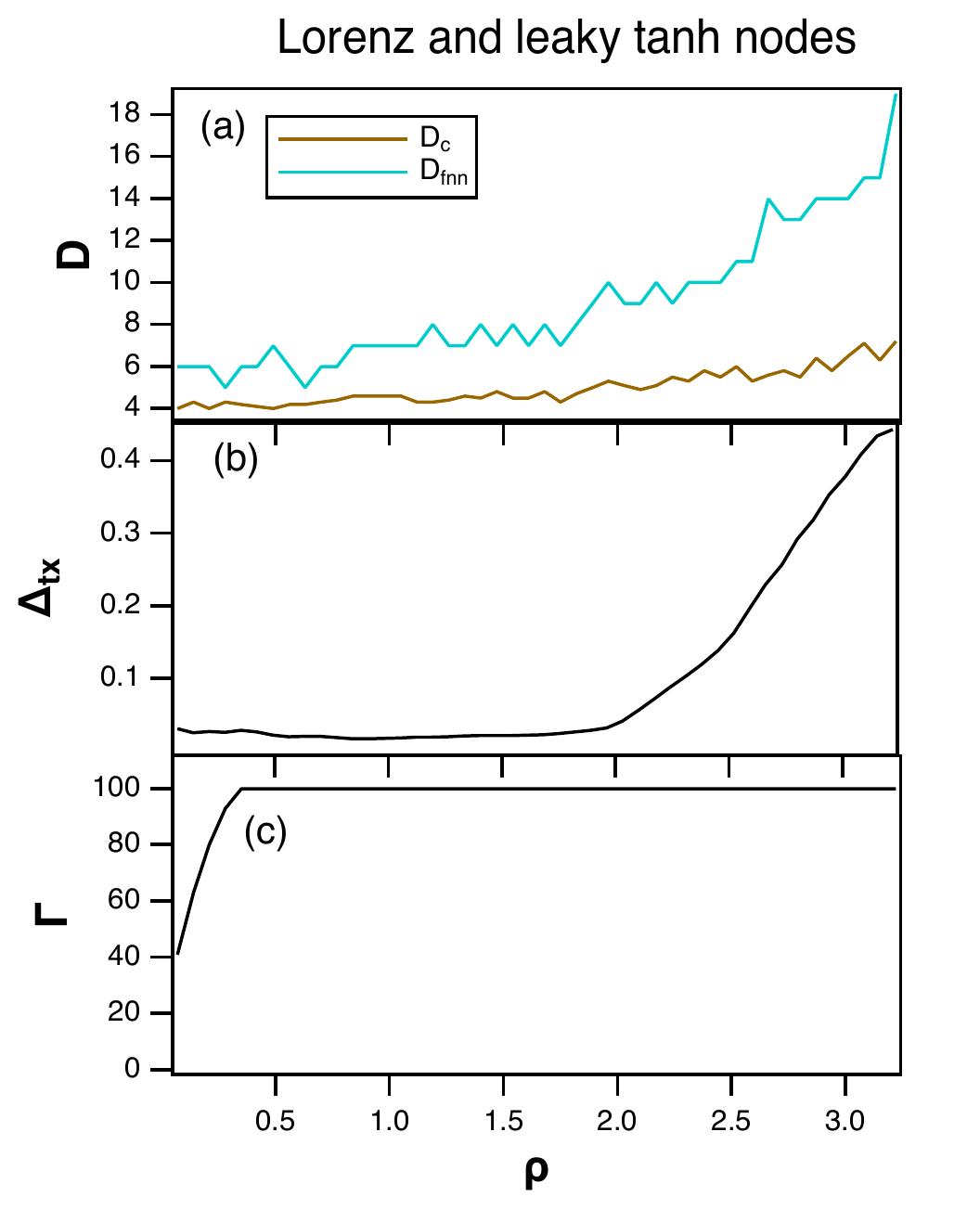} 
  \caption{ \label{lor_tanh_dim} (a) Covariance dimension $D_c$ and false nearest neighbor dimension $D_{fnn}$ of a reservoir consisting of 100 leaky tanh nodes driven by the Lorenz $x$ signal. As the spectral radius $\rho$ increases, both the covariance and false nearest neighbor dimensions increase. (b) Testing error $\Delta_{tx}$ for predicting the Lorenz $z$ signal from the Lorenz $x$ signal. (c) Covariance rank as defined in eq. (\ref{rank}).
}
  \end{figure} 

Figure \ref{lor_tanh_dim} shows the dimensions, testing error and rank for a reservoir of 100 leaky tanh nodes driven by a Lorenz $x$ signal. Similar to when the reservoir consisted of polynomial nodes (figure \ref{lor_poly_dim}), both covariance and false nearest neighbor dimensions increase as the spectral radius increases.  The testing error in figure \ref{lor_tanh_dim} also increases with dimension for $\rho > 2$. The plots only extend to $\rho = 3.2$ because the the largest Lyapunov exponent for the leaky tanh node reservoir driven by the Lorenz signal became positive at this value.

\begin{figure}[ht]
\centering
\includegraphics[scale=0.8]{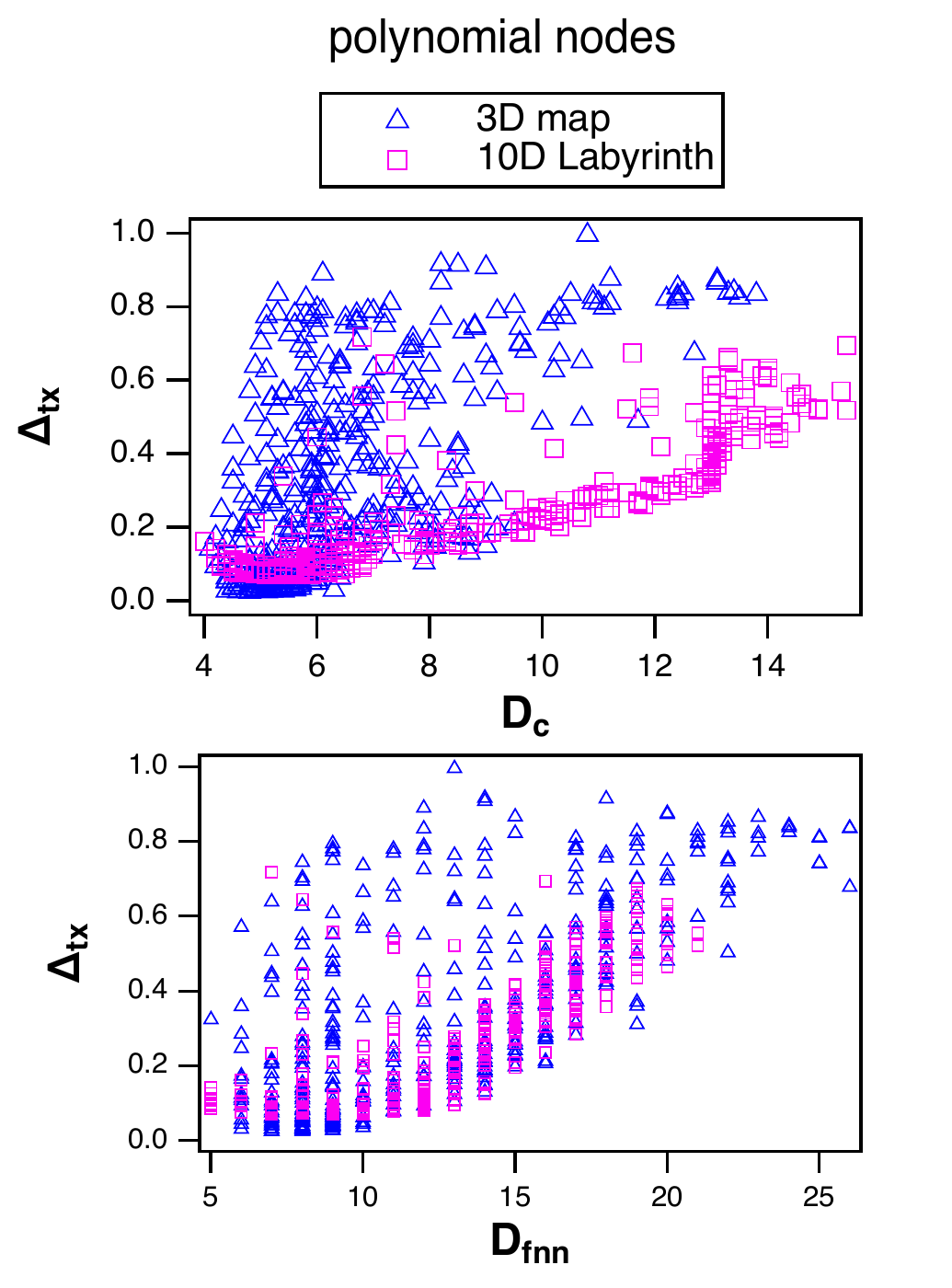} 
  \caption{ \label{nleq_dim_err} Top picture is the testing error $\Delta_tx$ for a reservoir computer using the polynomial nodes as a function of the covariance dimension $D_c$. The input signals came from the 3 dimensional map or the 10 dimensional Labyrinth system. The bottom plot is the testing error as a function of false nearest neighbor dimension $D_{fnn}$
}
  \end{figure}

\begin{figure}[ht]
\centering
\includegraphics[scale=0.8]{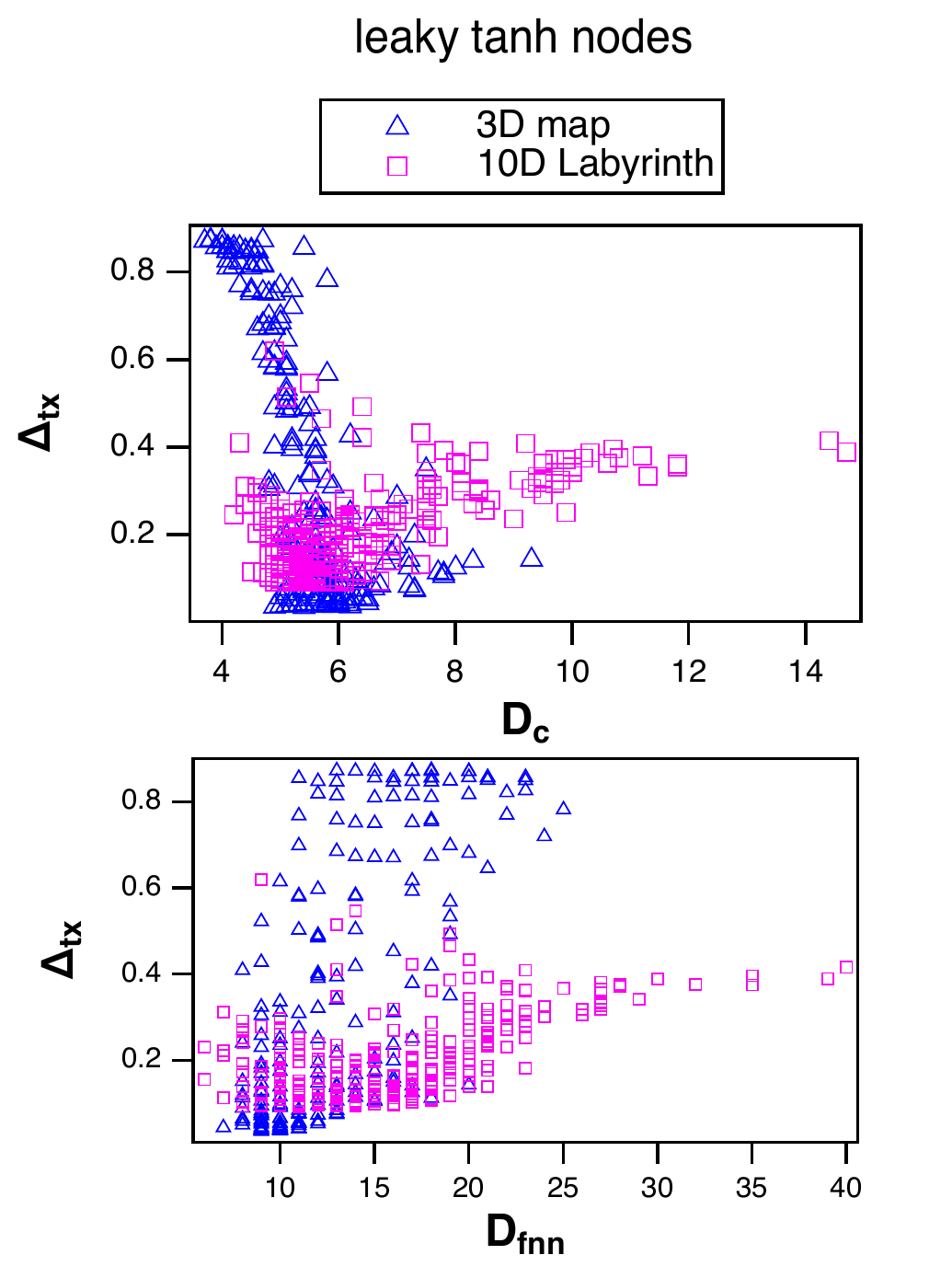} 
  \caption{ \label{umd_dim_err} Top picture is the testing error $\Delta_tx$ for a reservoir computer using the leaky tanh nodes as a function of the covariance dimension $D_c$. The input signals came from the 3 dimensional map or the 10 dimensional Labyrinth system. The bottom plot is the testing error as a function of false nearest neighbor dimension $D_{fnn}$
}
  \end{figure}

The relation between reservoir dimension and testing error for other combinations of signal and node is summarized in figures \ref{nleq_dim_err} and \ref{umd_dim_err}. Figure \ref{nleq_dim_err} shows the relation between reservoir dimension and testing error for the polynomial nodes when driven by the 3 dimensional map signal or the 10 dimensional Labyrinth signal. Both spectral radius and the parameter $p_1$ in eq. \ref{res_comp} were varied over their regions of stability.

There is a clear correlation between dimension and testing error for signals from the Labyrinth system in figure \ref{nleq_dim_err}, but the relation between dimension and testing error is less clear for the 3 dimensional map signal. The covariance and false nearest neighbor dimensions are the local dimensions of the manifold occupied by the reservoir; they are not the dimensions of the individual node signals $\chi_i(t)$ for the reservoir. The dimensions of signals $\chi_i(t)$ will be estimated in section \ref{dimchange}.

 Figure \ref{umd_dim_err} shows the relation between reservoir dimension and testing error for the leaky tanh nodes when driven by the 3 dimensional map signal or the 10 dimensional Labyrinth signal. The spectral radius and feedback parameter $\alpha$ in eq. \ref{umd_comp} were varied to create these plots.

As in figure \ref{nleq_dim_err}, there is a clear correlation between dimension and testing error for signals from the Labyrinth system in figure \ref{umd_dim_err}, but the covariance dimension for the leaky tanh nodes decreases as the covariance dimension increases, while there is no trend for the false nearest neighbor dimension. 

Again, the covariance and false nearest neighbor dimensions are the dimensions of the reservoir manifold. The figures in this section show that in some cases, there may be a relation between the manifold dimension and the testing error, but the manifold dimension may not be the most important quantity for a good reservoir computer. The dimension of the actual reservoir signals, $\chi_i(t)$, may be estimated from the Kaplan-Yorke dimension in section \ref{dimchange}. The trends with respect to Kaplan-Yorke dimension will be more clear.

\section{Dimension Increase due to Filtering Effects}
\label{dimchange}
Not only does the dimension of the manifold occupied by the set of the reservoir signals change, the fractal dimension of the individual reservoir signals can change because of a well known effect in driven dynamical systems \cite{badii1988}. The Lyapunov exponents of the driven system consist of a set of Lyapunov exponents from the driving signal plus a set of exponents due to the driven system. Although the Lyapunov exponents from the driven system are negative, they can replace some of the negative exponents from the driving signal in the summation used to calculate the Kaplan-Yorke dimension (eq. \ref{dimky}). These added exponents can cause a change in the fractal dimension in the signals in the driven system. If the added negative exponents are close enough to zero, they can even increase the integer part of the dimension. 

As an example, the Lorenz Lyapunov exponents for the parameters in eq. \ref{lorenz} are 0.9, 0 and -14.5. When the polynomial nodes were driven by the Lorenz $x$ signal, with the spectral radius for the reservoir $\rho=3$ and the constant $p_1=-3$ in eq. \ref{res_comp},  the largest Lyapunov exponent for the polynomial reservoir was -3.7.  The cumulative sum of the exponents for the Lorenz system alone is 0.9, 0.9, -13.6, so the Kaplan-Yorke dimension for the Lorenz system (eq. \ref{dimky}) is $2+(0.9+0.0)/14.5=2.06$. The cumulative sum of the exponents for combined system of Lorenz and polynomial reservoir was 0.9, 0.9, -2.8, -17.3, so the Kaplan-Yorke dimension is now $2+(0.9+0.0)/2.8 = 2.32$. 

\begin{figure*}[ht]
\centering
\includegraphics[scale=0.8]{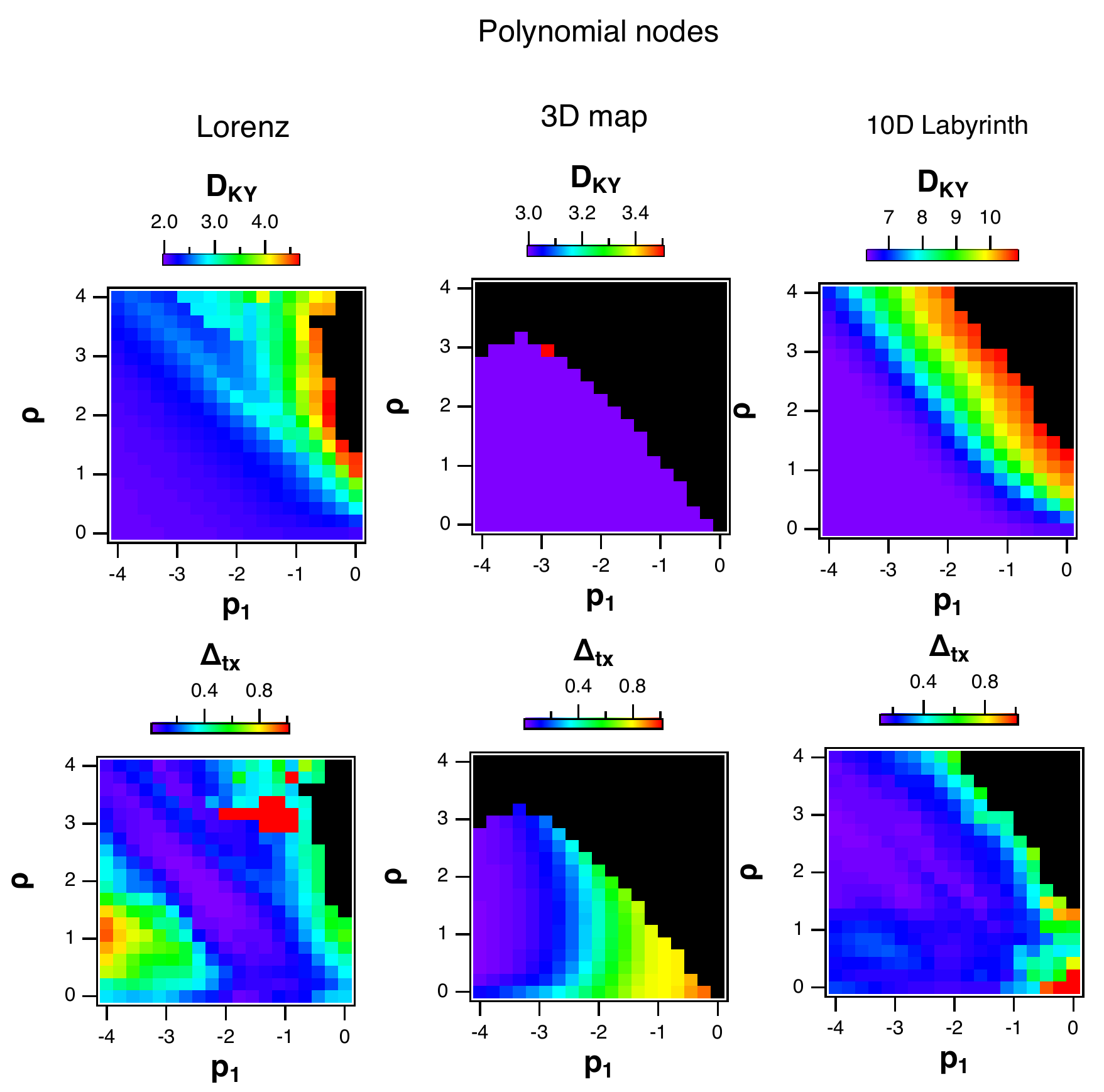} 
  \caption{ \label{nleq_dky_err} Results for the polynomial nodes. The top plots are the Kaplan-Yorke dimension $D_{KY}$ for the polynomial reservoir driven by the Lorenz $x$ signal, the 3d map signal, or the 10d Labyrinth signal, as the spectral radius $\rho$ and parameter $p_1$ vary. The bottom plots are the training error $\Delta_{tx}$ for the same three combinations. The black regions correspond to a positive Lyapunov exponent for the reservoir.}
  \end{figure*}

Figure \ref{nleq_dky_err} shows the Kaplan-Yorke dimension and the testing error for the polynomial reservoir driven by the three different input signals as $\rho$ and $p_1$ are varied.

The relation between the Kaplan-Yorke dimension and the testing error in figure \ref{nleq_dky_err} is complicated, but one feature is that the smallest testing errors occur at or near the lowest Kaplan-Yorke dimension. What this means is that if the fractal dimension of the reservoir signals $\chi_i(t)$ is significantly different from the fractal dimension of the training signal, then the testing error will be larger than the minimum. If the reservoir signals have a different fractal dimension than the signals from the driving system, the quality of the fits to the testing and training signals will be degraded.

\begin{figure*}[ht]
\centering
\includegraphics[scale=0.8]{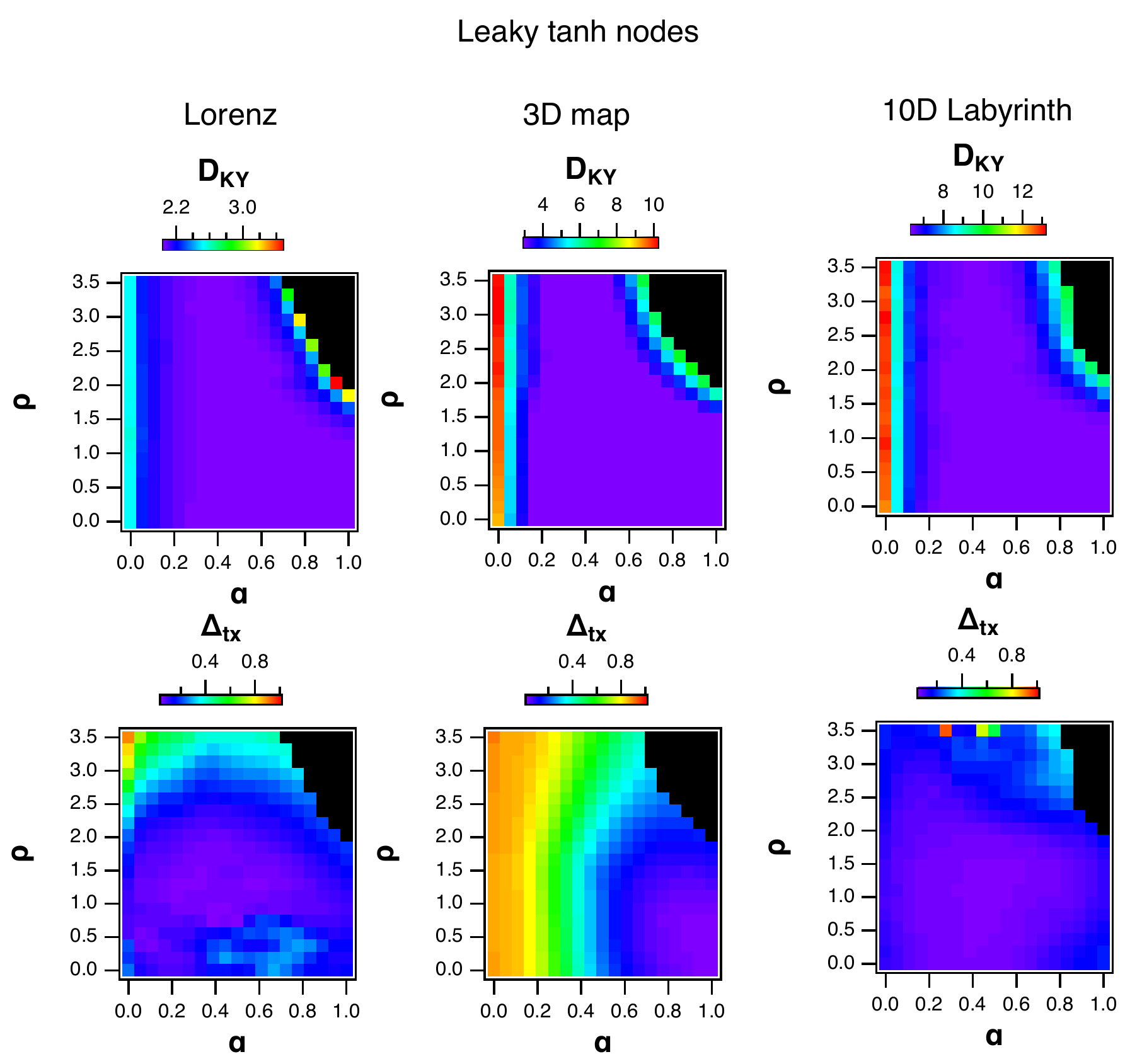} 
  \caption{ \label{umd_dky_err} Results for the leaky tanh nodes. The top plots are the Kaplan-Yorke dimension $D_{KY}$ for the polynomial reservoir driven by the Lorenz $x$ signal, the 3d map signal, or the 10d Labyrinth signal, as the spectral radius $\rho$ and parameter $p_1$ vary. The bottom plots are the training error $\Delta_{tx}$ for the same three combinations. The black regions correspond to a positive Lyapunov exponent for the reservoir.}
  \end{figure*}

The Kaplan-Yorke dimension and testing error for the leaky tanh nodes are shown in figure \ref{umd_dky_err}.

Figure \ref{umd_dky_err} also shows that the regions of the smallest testing error in the different reservoirs correspond to the regions of smallest Kaplan-Yorke dimension. A small Kaplan-Yorke dimension by itself is not sufficient to guarantee a small testing error, but a small Kaplan-Yorke dimension is necessary. It may also be noted in figures \ref{nleq_dky_err} and \ref{umd_dky_err} that the smallest training error does not always occur at the edge of chaos. Regions of instability, indicated by a positive Lyapunov exponent for the reservoir, are shown in black in figures \ref{nleq_dky_err} and \ref{umd_dky_err}.

\section{Edge of Chaos}
There are many examples in the literature that suggest the best parameter region for operating a reservoir is at the edge of the stable region, or the "edge of chaos" \cite{lukosevicius2012}, although some have argued  that this edge of chaos rule is not universal \cite{mitchell1993,mitchell1993a}.  Figures \ref{nleq_dky_err} and \ref{umd_dky_err} show that in some cases, the smallest training error is not at the edge of chaos. As the Lyapunov exponents of the reservoir increase towards zero, they may overlap with the Lyapunov exponents of the driving signal, increasing the fractal dimension of the reservoir signals to the point that they do not provide a good fit to the training or testing signals.

The idea that a dynamical system or a cellular automaton has its greatest computational capacity at the edge of chaos was introduced in \cite{packard1988,langton1990,crutchfield1990}. The edge of chaos is a phase transition between an ordered state and a disordered state. Systems in the vicinity of this phase transition exhibit the most complex behavior of any parameter range, and thus have the greatest capacity for computation. Because dynamical systems have the greatest computational capacity at the edge of chaos, it is believed that a reservoir computer will function best at this edge.

The reason that the parameters where a reservoir computer performs optimally may not be the same parameters that give the maximum computational capacity is that reservoir computers as applied in this work are not being used as general purpose computers. Rather, in this work the reservoir is being used to learn the nonlinear function between an input signal and a training signal. In \cite{carroll2019} showed, this function approximation is similar to fitting a signal with a set of orthogonal basis functions; the higher the rank of this basis (the covariance rank in this paper), the better the fit. In addition, the reservoir must be synchronized to the input signal in the general sense.  The continuity between the reservoir and the input signal was measured in \cite{carroll2018}, where it was demonstrated that higher the value of a continuity statistic  between the reservoir and the input signal, the lower the training error.

 As figures \ref{nleq_dky_err} and \ref{umd_dky_err} show, if the fractal dimension of the reservoir signals change significantly from the fractal dimension of the input signal, the testing errors become larger. If the reservoir fractal dimension is too large, then there may be no continuous function between the reservoir and the input system, so that generalized synchronization is not possible.
 
 What is shown in this work is that matching the reservoir dynamics to the input signal is important. Much is made of the memory capacity of the reservoir \cite{jaeger2001}. Memory is a direct result of the Lyapunov exponent spectrum of the reservoir. The closer the largest Lyapunov exponent for the reservoir comes to the edge of stability, the longer the memory. The work here demonstrates that memory must be matched to the task at hand; a memory that is too long will mix together parts of the signal that are uncorrelated with each other, destroying generalized synchronization, while a memory that is too short may not reproduce all the relevant time scales in the input signal.

\section{Summary}
It was shown that overlap between the Lyapunov exponent spectrum of a reservoir driven by a chaotic signal and the Lyapunov exponents of the chaotic driving system could mean that the signals in the reservoir, $\chi_i(t)$, had a larger dimension than the driving system. When the reservoir signals had a higher dimension than the driving system, the testing error was larger than when the reservoir and driving system had the same dimension. This dimension difference points out that for best performance it is necessary to match the reservoir to the signals being studied. If the reservoir has a higher dimension than the driving system, then there is no continuous function from the reservoir to the driving system, hence no generalized synchronization.

It is common to say that a reservoir computer works best at the "edge of chaos", but this claim was not true for the signals and reservoirs in this work. As the largest Lyapunov exponent of the driven reservoir approaches zero from below, it will replace some of the driving system Lyapunov exponents in the calculation of the Kaplan-Yorke dimension, leading to a mismatch between the fractal dimension of the driving system and the reservoir. It was shown that this mismatch leads to an increased testing error.

It was also shown that the reservoir signals exist on a low dimensional manifold. The dimension of this manifold itself did not have a large effect on the reservoir computer performance, but the manifold dimension was sometimes correlated with reservoir computer performance because of changes in the reservoir fractal dimension.

This work was partially supported by the Office of Naval Research through the Science of AI program and the Naval Research Laboratory's Basic Research Program.

\bibliography{reservoir_dim.bib}{}

\end{document}